# Characterization of plasma and gas-phase chemistry during boron-nitride nanomaterial synthesis by laser-ablation of boron-rich targets

Shurik Yatom and Yevgeny Raitses

Princeton Plasma Physics Laboratory, Princeton University, NJ, USA.

## Abstract

In this work, solid targets made from boron and boron nitride (BN) materials are ablated by a nanosecond pulsed laser at sub-atmospheric pressures of nitrogen and helium gases. Excited species in the ablation plume from the target are probed with spatiotemporally resolved optical emission spectroscopy (OES). Evaluation of chemical composition in the plasma plume revealed that for both boron-rich targets, emission from BN molecules is always observed in nitrogen-rich environments. In addition, BN molecules also present when ablating a boron nitride target in helium gas- an indication that BN molecules in the plume may originate from the solid target. Furthermore, the ablation of BN target features emission of $B_2N$ molecules, regardless of the pressure and surrounding gas. These results suggest that the ablation of the BN target is more favorable for the generation of complex molecules containing boron and nitrogen species and possibly hint that BN is also more favorable feedstock for high-yield BN nanomaterial synthesis. Plasma parameters such as electron temperature (peak value of 1.3 eV) and density (peak value of $2\times10^{18}$ cm$^{-3}$) were also investigated in this work in order to discuss the chemical dynamics in the plume.

## 1. Introduction

Boron-nitride (BN) nanomaterials have attracted a great interest due to their unique electrical, thermal and mechanical properties. Boron-nitride nanotubes (BNNTs) have been predicted in 1994[1] and have been synthesized in 1995.[2] Since their discovery BNNTs invoke broad inter-disciplinary interest due to their vast potential in a wide variety of applications. BNNTs stiffness and strength allow manufacturing of strong and lightweight composites that are capable of enduring high temperatures.[3,4] BNNTs are electrical insulators with high thermal stability, therefore they are expected to be useful for thermal management in, for example, electronics,[5] aerospace etc. Studies showed that BN materials can be used in as efficient, non-toxic catalysts.[6,7] Optics and photonics are another area where BNNTs have a potential as light-emitting diodes and photodetectors with narrow wavelength selectivity for emission and detection.[8,9] Other areas of potential applications for BNNTs include: vibrational damping,[10,11] piezo-electric[12] and hydrophobic materials,[13] neutron sensing[14] and shielding,[15] and applications in oncology[16] and drug delivery.[17]

The synthesis of BNNTs has been successfully demonstrated by a variety of techniques such as: atmospheric pressure an high pressures arcs,[18,19] laser ablation[20,21,22] and inductively-coupled plasma torches.[23,24,25] However, physico-chemical mechanisms by which the BNNTs are synthesized are not well understood. One possible mechanism is the so-called "root-growth" - was proposed in the work of Arenal et al[20] and further developed in Ref. [26]. The principle of "root-growth" mechanism requires the formation of a molten boron seed particle - boron droplet at the temperature above 2000 K, which is bombarded by nitrogen atoms and nitrogen-containing molecules, thereby sustaining the BNNT growth from the droplet. This mechanism received evidences in post-run characterization of synthesized products which revealed BNNTs connected to boron droplets.[18,24] Other experiments and molecular dynamic (MD) simulations suggest that boron droplets are not necessary for the facilitation of BNNT synthesis in plasma[27]. Independent on the primary mechanism of the growth, their validation requires in situ characterization of synthesis processes.[28,29,30]

During the plasma-based synthesis, the main plasma processes are ionization, dissociation, association, and recombination. Since plasma promotes various plasma-chemical reactions and processes, it is critically important to identify the gas phase/plasma phase precursors sustaining the growth. In molecular dynamics simulations, synthesis of BNNTs and BN nanostructures was explored by means of boron droplet (at the temperature of 2000-2400 K) bombardment by nitrogen atoms,[26] BN molecules,[31] synthesis from molecular precursors such as borazine ($B_3N_3H_6$) and iminoborane (HBNH),[31,32] radical BN[32] with/without addition of hydrogen and mixtures of B/N/H in different ratios.[32] In experiments, the chemistry induced in plasmas that contain B and N species was probed by optical emission spectroscopy (OES) to survey the emission of species in laser ablation[33] and induction thermal plasma.[34] OES results detected mainly BN, $N_2$, $N_2^+$, BH and NH molecules and discussing their possible roles in synthesis of BN films or nanotubes. Infrared spectroscopy of laser ablation of B target in $NH_3$ showed formation of complex species, including $B_2N$, iminoborane and borazine[35].

In a recent study of the gas composition for high temperature growth of BN nanomaterials,[36] the formation of different chemical species relies on using the thermodynamic approach of minimization of the Gibbs free energy.[36,37] The results showed that for B-N and B-N-H gas mixtures under thermodynamic equilibrium at pressures 1- 10 atm and gas temperatures of $T_g$<4000 K, $B_2N$ molecules dominate and their density outnumber BN molecules by 2-3 orders of magnitude.[36] In addition, this work presented also the first experimental documentation of the emission by excited $B_2N$ molecules in the visible spectrum, which was recorded in the laser ablation plume from a solid BN target. In the present paper, we are going beyond these initial studies of laser ablation of solid BN targets and focus on a comparison of gas phase reactions in the ablation plumes from boron and boron nitride targets. Here, we report spatiotemporally resolved investigation of atomic and molecular emission of the excited species in the ablation spectra of these solid targets in different gases such as $N_2$, $N_2$ with admixed $H_2$ and He. In these experiments, the gas pressure was at 500 Torr. In addition, we also conducted experiments on ablation at much lower pressure of 100 mTorr. Our experiments revealed differences in the measured spectra of the ablation plumes between the boron and BN targets. These results indicate on possible differences in the ablation mechanisms of these materials. In particular, it appears that BN solid target favors formation of large molecules such as $B_2N$. In order to support the analysis of emission dynamics we also measured the ablation rate of boron and boron-nitride targets and the characterization of plasma in the ablation plumes, via investigation of atomic line broadening and intensity of different excitation stages of hydrogen atom.

## 2.Experimental setup and procedure

The experimental setup for the ablation consists of a stainless steel 6-way cross equipped with four viewports allowing laser beam access through the window, into the chamber to the ablated target, and a view of the ablation plume in direction normal to the laser beam. The boron (rod-shaped ingot: 1 cm diameter, 2 cm height, purity 99.5 %) and boron-nitride (disc: 2.5 cm diameters, 5 mm thickness, purity 99.5 %) targets are situated on 3D printed adapter tray, which is connected to a small, rotating electric motor. The ablation is driven by nanosecond laser pulses, provided by Nd: YAG Continuum Surelite III laser. The fundamental harmonic of the laser frequency-tripled by a BBO crystal, generating 355 nm laser pulses with full-width-at-half-maximum (FWHM) duration of ~5 ns and energy ~150 mJ. The laser beam is focused on the target with a lens, down to a ~200 µm spot, therefore the laser fluence is evaluated at ~120 J/cm$^2$ (irradiance ~15 GW/cm$^2$). Laser pulses are shot at frequency of 10 Hz at the rotating target. The ablation plume is formed at the point of laser impact and extends in the horizontal direction (co-axial to the laser beam direction). The plume is imaged on the entrance slit of the imaging spectrometer (iHoriba 550). An ICCD camera (PIMax 3 by Princeton Instruments) is attached to the spectrometer, serving as a 2D signal detector. The plume image is rotated $90^0$ by a Dove prism so that the plume length is imaged across the vertical direction of the ICCD. In this way, we have an ICCD image, where the horizontal axis corresponds to dispersion (represent the wavelength) and the perpendicular axis correspond to the spatial distance in direction normal to the target. A BNC digital pulse delay generator was used to synchronize the operation of the laser with the iCCD camera. The initial exposure of the iCCD is timed to coincide with the end of the laser pulse. This is in order to avoid the saturation of the iCCD by the elastically scattered laser light and the strong continuum radiation right after the laser impact on the ablation target.

The emergence of the first spectral lines was observed with the spectrometer at approximately 10 ns after the end of the laser pulse. The data presented in this paper uses the beginning of the observable laser pulse as the reference time (t=0). The temporal evolution of the spectra was captured while increasing the delay of the iCCD camera acquisition up to few microseconds with respect to the laser shot. The intensity of the emission from different excited elements varied strongly between different species and at different delay times, therefore capturing emission of various elements required multiple accumulations at the same time delay. The exposures and the amount accumulations vary between 20-500 ns for exposure and 50-800 accumulations. The approach of using accumulations relies on assumption that the temporal evolution of chemical kinetics and plasma parameters are similar between the subsequent laser shots.

The spectra were obtained in the following two main configurations of the spectrometer. Spectral survey of the elements present in the discharge and the study of the emission intensity of various elements were conducted using the low resolution (150 gr/mm) grating of the iHoriba spectrometer. To facilitate the analysis of plasma electron density we have added hydrogen (10 % of total gas composition) to nitrogen gas. The line broadening of hydrogen Balmer α ($H_\alpha$) line was measured using the low resolution grating in the first stages of the discharge, when the line profile was very broad. In the later stages of the discharge the line broadening of $H_\alpha$ line was studied using the higher resolution grating (1200 gr/mm). The spectral resolution and the instrumental broadening corresponding to the two gratings were measured using the calibration lamp (Oriel HgAr pencil lamp) and were determined to be $\Delta_{instr}$~0.7 nm for 150 gr/mm grating and $\Delta_{instr}$=0.079-0.038 nm for 1200 gr/mm grating. Finally, the transmission of the optical system for different spectral regions was calibrated with an integrated sphere light source

(Labsphere URS-600).

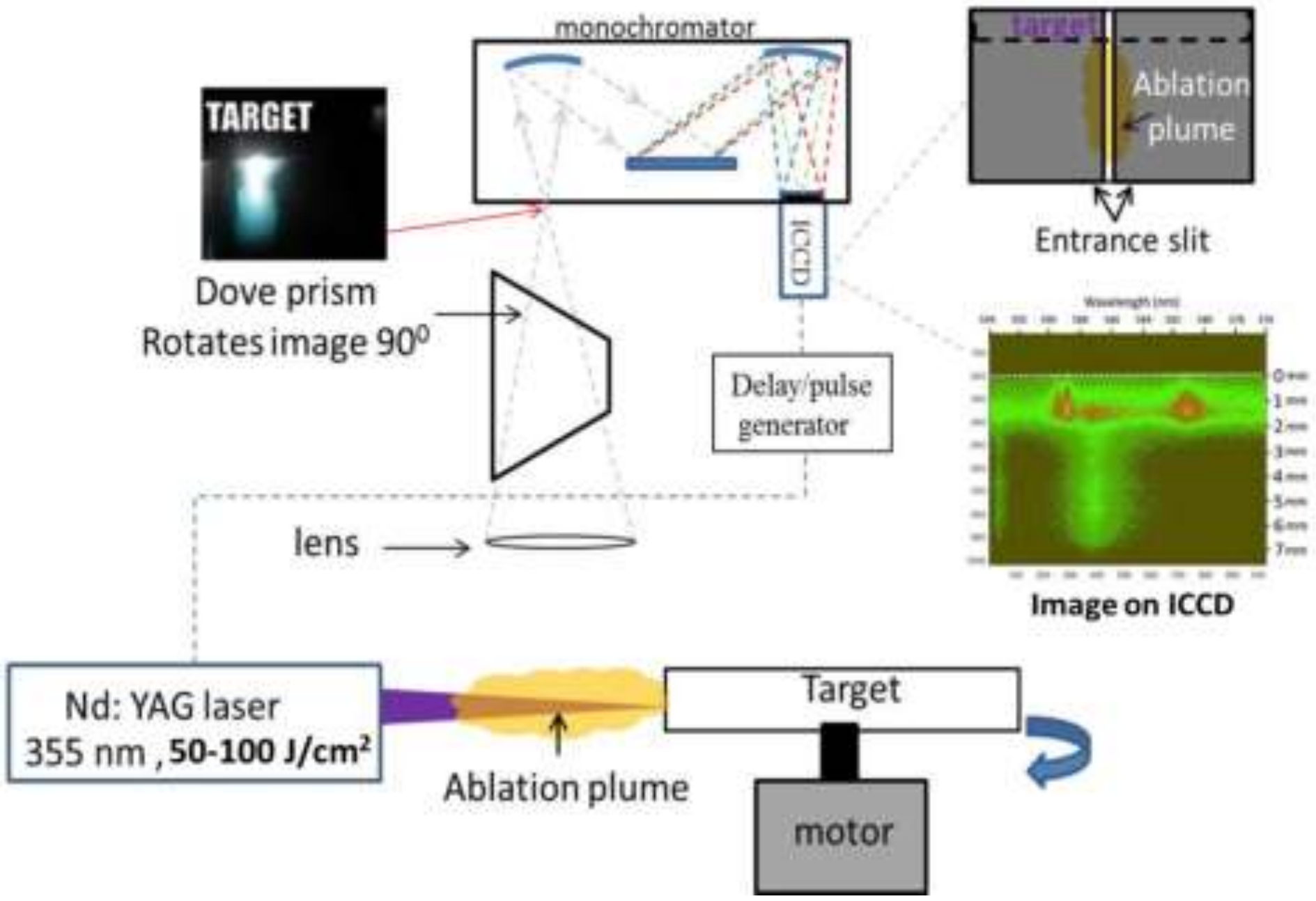

Figure 1. Experimental setup for optical emission spectroscopy study of laser ablation of B and BN targets.

## 3. Experimental results

3.1 Spectral survey and evolution of emission

We have examined the emission of the ablation plume in the visible spectrum (330-900 nm) during ablation of both boron and solid BN targets in low (100 mTorr) and high (500 Torr) pressures of helium and nitrogen gases. The identification of atomic lines relies on NIST database.[38] Emission of BN (0,0) band was detected in ablation of both BN and B targets (Figure 2). BN emission was recorded in ablation of B and BN targets in 500 Torr of $N_2$. In addition, BN emission was also observed when the BN target was ablated in 100 mTorr pressures. For both low- and high-pressure regimes, the ablation spectrum of the BN target also shows unidentified strong spectral features at ~362.15, 362.8 and 364.2 nm. These features were also present in the results shown in the work of Dutouquet et al.[39] We hypothesized that these are molecular band-heads, due to the fact that these features stay visible up to 4 µs after the laser shot (at P=500 Torr). The emission of $B_2N$ molecules was also observed. The identification of two $B_2N$ band-heads was availed due to the calculated vibronic spectra of $B_2N$, presented in Ref. 40. The evolution of the emission features in the spectral window shown in Fig. 3 includes both of these bandheads and the emission of several N II lines at 500-500.6 nm. The observed bandheads match the $\Sigma^+(0,1,0) - (0,1^1,0)$ and $\Delta^1(1,1,0) - (0,1^1,0)$ transitions of $^{11}B^{14}N^{11}B$ as postulated in Ref. 40. The emission of NII in this spectral window serves as the reference for the identification of the $B_2N$ bands. Finally, a high-resolution spectra was taken in the 470-530 nm region in order to detect other $B_2N$ bandheads which were, for example, calculated by Ding et al.[40] The obtained results include various features with emission persisting for t > 4 μs (at P=500 Torr), but none of these results matched the rest of $B_2N$ bandheads indicated in Ref.40.

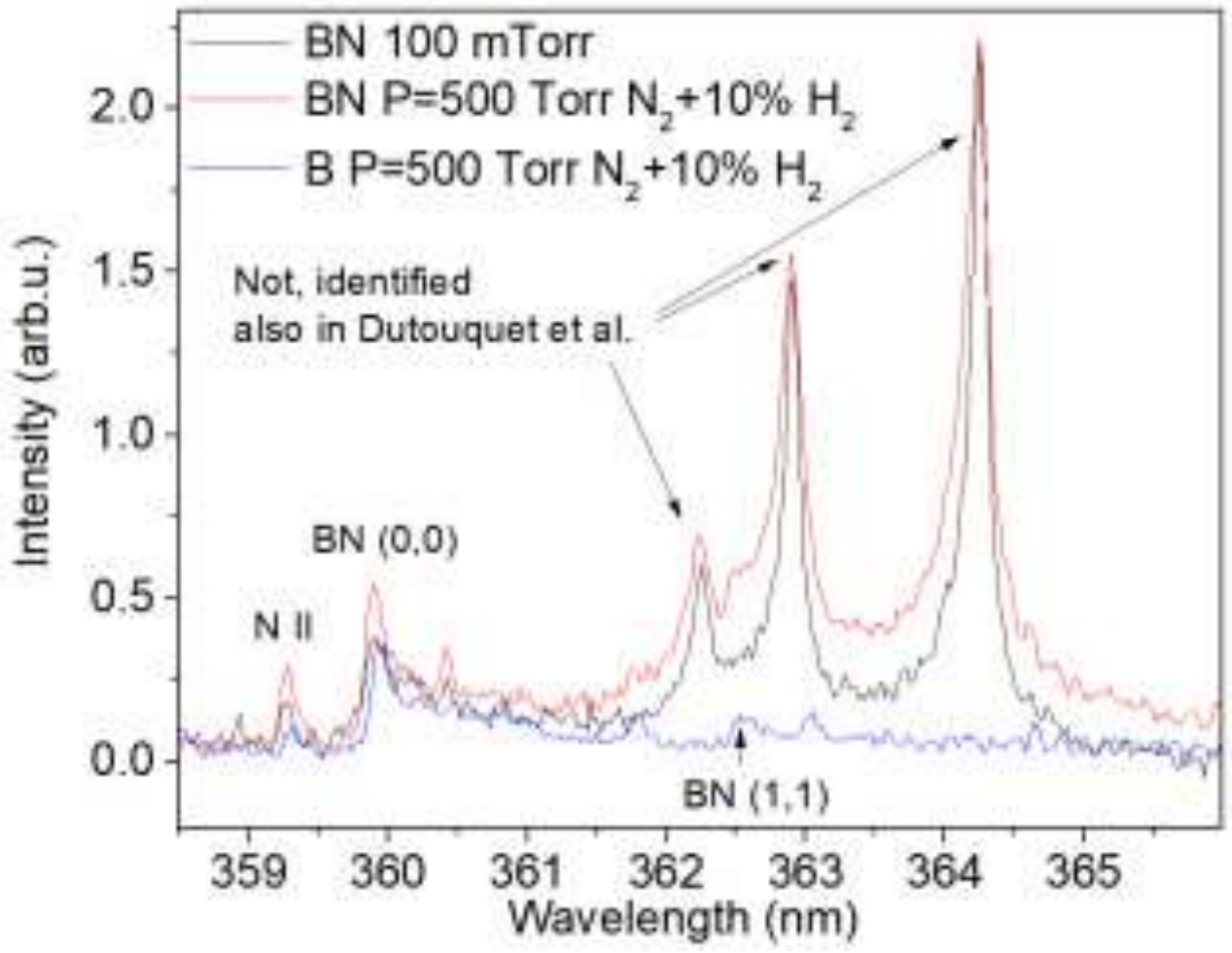

Figure 2. Emission of BN molecule recorded in ablation of B and BN targets. Different color curves correspond to different experimental conditions (ablated target and pressure). The emission intensities were normalized to fit a single plot, therefore they only reflect the relative emission intensity evolution as a function of time and are not to be compared between different species.

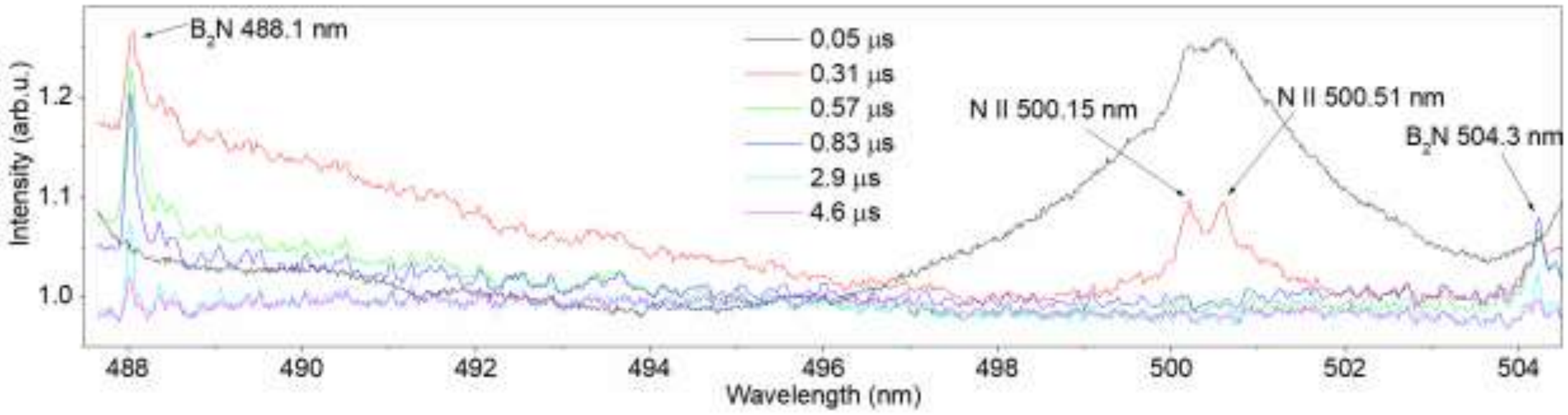

Figure 3. Emission of $B_2N$ molecular bands. Different color curves correspond to spectra captured at different times with respect to the laser shot. Each curve recorded at the same conditions and were not normalized, hence the signals represent the actual intensity of emission by the different species .

One of the interesting observations is that the ablation of BN target showed both the BN and $B_2N$ emission independent on the background gas and pressure, i.e. in vacuum (100 mTorr), helium and nitrogen (P=500 Torr). In contrast, the spectra measured in the ablation plume from the boron target features only BN (0,0) at 359.9 nm, while $B_2N$ bandheads were not detected. This is despite the fact that these measurements were conducted with long exposures up to 400 ns and accumulations of up to 600 pulses.

Figure 4 shows the temporal evolution of emission intensity from different atomic and molecular species in the ablation plume from BN and B targets. Here, the intensities were normalized to be shown together in the same plot. They represent the temporal evolution of emission by different species, but do not reflect the differences in their emission intensity. Figure 4a shows the temporal evolution of emisison in the ablation plume from BN target in vacuum (P=100 mTorr). In this case, the target serves as the source of generated species in the plume. The emisison spectra of atomic species were recorded with the exposure of 20 ns, whereas the emission spectra of BN and $B_2N$ molecules were recorded with the exposure of 50 ns.  Under such conditions, the onset of emisison by BN and $B_2N$ molecules appears very early, at t=50 ns, which is also when the maximum intensity of BN(0,0) emission is obtained. The peak of $B_2N$ is slightly delayed, appearing at t=110 ns. The emisison from B and N ions also appears early at t=50 ns and peak before 100 ns. The emisison of neutrals onsets later with respect to ions and molecules , reaching their peak at t=200 ns.  . The times of first appeareance and peak intesity of ions, neutrals and molecules are summarized in Table.1 for the different targets and pressures.

Another interesting observation shown in Fig. 4a is the apparent incresase in the intentisy of BN(0,0) emission at 200<t<500 ns, resulting in a local maximum around t~500-600 ns (indicated as stage III on the plot). The early emergence of BN and $B_2N$ emisison is not exclusive to BN ablation in vaccum and also observed in BN ablation in helium at P=500 Torr. Figure 4c compares BN and $B_2N$ intensities for the ablation of BN target in vacum (100 mTorr), helium, and nitrogen at P=500 Torr. One can see that the helium and vacuum cases exhibit very similar evolution for BN and $B_2N$ emission, peaking at t=50 ns and t=100 ns, respectively. BN ablation in nitrogen is quite different, because both molecules reach peak intensity at much later stage, t~ 550 ns. Figure 4b and 4d depict the emission evolution following the ablation of boron (4b) and BN target (4d) in nitrogen, where the emisison of the ions was omitted to render the plots more readable. Note that in the case of boron target (Figure 4b), the signal was weak and observable only with minimum exposure of 100 ns. It is evident that the temporal evolution of the emission of B, N and BN following the ablation of both targets is very similar, except the fact that no $B_2N$ emission was detected in the ablation plume of the boron target

Table 1. Timing of first appearance ($t_0$) and peak of intensity ($t_{peak}$) for ions, neutrals and molecular species in ablation plume of B and BN targets.

| **Element/** | B I | B II | N I | N II | BN | $B_2N$ |
|---|---|---|---|---|---|---|

| **Target material, pressure** | time(ns) | | time(ns) | | time(ns) | | time(ns) | | time(ns) | | time(ns) | |
|---|---|---|---|---|---|---|---|---|---|---|---|---|
| | $t_0$ | $t_{peak}$ | $t_0$ | $t_{peak}$ | $t_0$ | $t_{peak}$ | $t_0$ | $t_{peak}$ | $t_0$ | $t_{peak}$ | $t_0$ | $t_{peak}$ |
| B , 500 Torr | 40 | 180 | 30 | 100 | 40 | 160 | 40 | 100 | 180 | 750 | Not found | |
| BN, 500 Torr | 40 | 180 | 30 | 100 | 40 | 150 | 30 | 100 | 115 | 600 | 120 | 800 |
| BN, 100 mTorr | 100 | 250 | 50 | 100 | 100 | 250 | 50 | 100 | 50 | 50 and 500 | 50 | 110 |

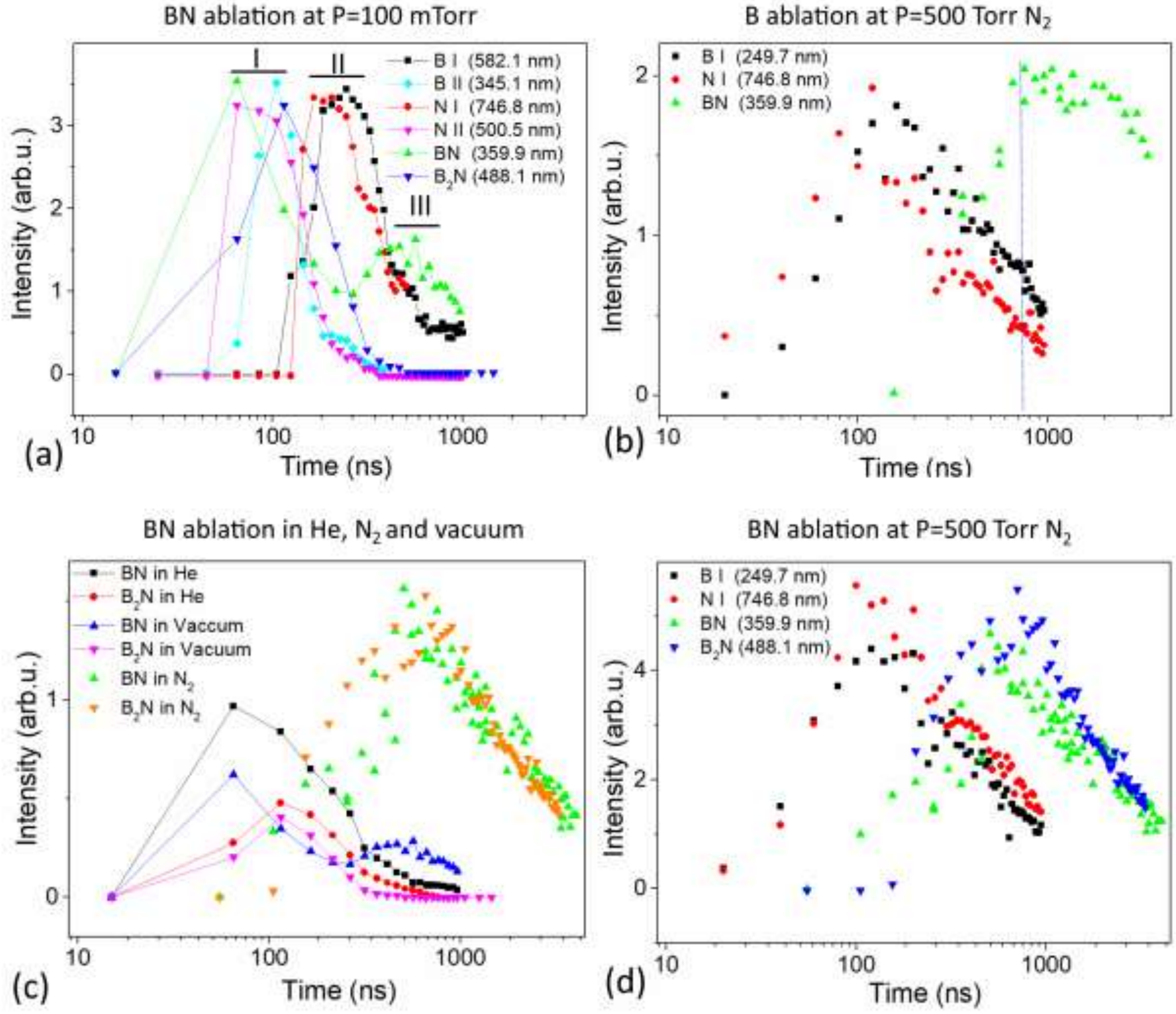

Figure 4. Temporal evolution of the emission by different species. (a) Emission of atomic and molecular species following BN target ablation in vacuum (P=100 mTorr) , phase I: peak emission of ions and molecules, phase II: peak emission of neutrals, phase III: secondary peak of BN emission. (b) emission of atomic and molecular species following B target ablation in 500 Torr of $N_2$, the blue vertical line is added to indicated the apparent peak of BN emission. (c) Comparison of BN and $B_2N$ emission following BN target ablation in He, vacuum (100 mTorr) and $N_2$. (d) Emission of atomic and molecular species following BN target ablation in 500 Torr of $N_2$. The emission intensities were normalized to fit a single plot, therefore they only reflect the relative emission intensity evolution as a function of time and are not to be compared between different species.

Figure 5 shows the spatially resolved spectra, including atomic species and $B_2N$ bandheads, along the axis perpendiculat to the surface of the target. Spatially, the emission of all elements, including molecules, starts adjacent to the target. The intensity is strongest at the vicinity of the target and extends to ~3-4 mm away, during first 200 ns (see Fig. 5a). Fig.5b shows two $B_2N$ bandheads at t=1000 ns, extending up to 7 mm. Here, the emission within the first mm of the target is gone. Our interpreation of these images is as the indication where the atomic species emitted by the target (and/or produced as a result of dissociation of molecules) participate in the reactions that produce the $B_2N$ molecules in the volume bounded by the first ~3 mm adjacent to the target. In the later stages, the molecules diffuse further away from the target, reaching 7 mm away from the target. In Appendix B, we also present a collage of images that shows the spatial evolution of B and N ions lines in the first stages of abaltion ( 0<t<200 ns).

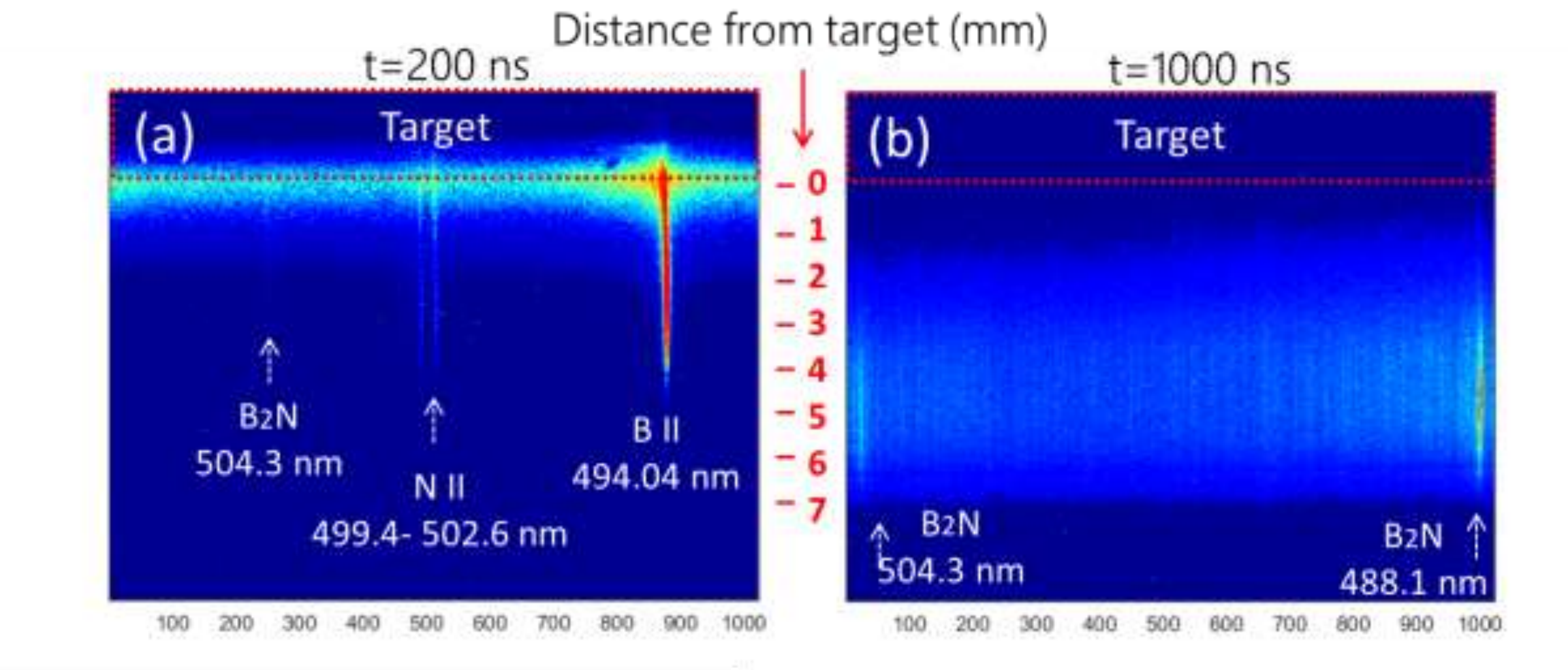

Figure 5. ICCD images showing spectra captured in ablation of BN target withspatial resolution in the direction "away form the target": (a) t=200 ns, ionic lines and $B_2N$ bandhead (b) t=1000 ns, $B_2N$ bandheads.

3.2 Analysis of line intensities

Temporal evolution of intensities of the Balmer α and β lines was examined in order to probe the electron temperature $T_e$. The ratio of the line intensities, with wavelengths $\lambda_{p'q'}$ and $\lambda_{pq}$ is given by[41]:

$$\frac{I(p\to q)}{I(p'\to q')} = \frac{\lambda_{p'q'}}{\lambda_{pq}} \frac{A_z(p\to q)}{A_z(p'\to q')} \frac{g(p)}{g(p')} exp\left(-\frac{E(p)-E(p')}{k_B T_{ex}}\right) \quad (1)$$

where $k_B$ is the Boltzmann constant, $A$ is the Einstein coefficient of the transition is, $g$ is the degeneracy of the level and $E$ is the energy of the upper level in the transition, $I(p \to q)$ and $I(p' \to q')$ are the intensities of the lines, corresponding to de-excitation from upper levels indexed by $p, p'$ to lower levels indexed by $q, q'$, respectively. The excitation temperature $T_{ex}$ is defined for a population of particles with Boltzmann distribution. Because of the exponential nature of this dependence, it is most sensitive when $k_B T_{ex} < E(p) - E(p')$. The upper level energy difference between $H_\alpha$ and $H_\beta$ is ~0.66 eV, therefore that pair is best utilized for plasmas with $T_{ex}$ <0.66 eV. Under the assumption of thermodynamic equilibrium, the excitation temperature $T_{ex}$ is equivalent to electron temperature $T_e$. Then, for the electron density, following Griem,[42] the rate of electron collisions with the given species having the largest energy gap $[E(p) - E(q)]$ should be larger than radiative decay at least by a factor of 10:

$$\frac{n_e}{cm^{-3}} \geq 10^{14} \times \sqrt{\left(\frac{k_B T_e}{eV}\right)} \left[\frac{E(p) - E(q)}{eV}\right]^3 \ (2)$$

Since, our analysis of Eqs. (1) and (2) relies on the intensity of the lines; the opacity effect was evaluated using the following relationship: [42]

$$\tau(\lambda_0) = \pi r_e \lambda_0 f_{ik} n_i d \sqrt{\frac{Mc^2}{2\pi k_B T_a}} \ (3) ,$$

where $T_a$ is the temperature of the absorbing atoms, $r_e$ ($2.82 \times 10^{-13}$ cm) is the classical electron radius, $f_{ik}$ (0.641 and 0.119 for $H_\alpha$ and $H_\beta$ respectively ) is the absorption oscillator strength, $\lambda_0$ is the transition wavelength (656.3 and 486.1 nm for $H_\alpha$ and $H_\beta$ respectively), $d$ =0.5 mm is the length of the absorbing homogeneous plasma slab, $M$ is the atom mass ($1.67355 \times 10^{-24}$ g for hydrogen), $n_i$ is the population density of the lower level, and $c$ is the speed of light. The population density of the lower level in the transition is calculated using Boltzmann statistics, i.e. $n_i = N_a exp\left(-\frac{E_i}{k_B T_a}\right)$, where density of absorbers $N_a$ is calculated through a hydrogen atoms percentage in the total gas mixture, assuming full dissociation of $H_2$ (for evaluation of the "worst case scenario", i.e. maximum amount of potential absorbers). The upper limit of the population is estimated for $T_a \approx T_e \sim 1$ eV and yields values $\tau(\lambda_0) \sim 10^{-5} - 10^{-6} \ll 1$, for $H_\alpha$ and $H_\beta$, safely rendering them as optically thin.

3.3 Analysis of line broadening

The spectra obtained in laser ablation experiments featured both $H_\alpha$ and $H_\beta$ lines (see Appendix B). For the determination of the plasma density, $H_\alpha$ is more appropriate as its higher intensity gives a better signal-to-noise ratio (SNR).The emission of $H_\alpha$ appears ~20 ns earlier and lasts longer than that of $H_\beta$. Moreover, the $H_\beta$ is much broader than $H_\alpha$ at the same plasma density ($N_e$) and is hard to process. Although Balmer α line is also very broad in the first few hundreds ns, it is still possible to process and fit $H_\alpha$ profiles.

Figure 6 shows the evolution of the $H_\alpha$ line at times $80 \leq t \leq 400$ ns, through several selected spectra recorded in that time interval during the ablation of a boron target. Spectral line broadening is caused by several mechanisms, such as instrumental broadening due to finite resolution capability of the optical setup, Doppler broadening due to thermal motions of emitters, broadening due to collisions with neutral particles (Van der Waals and resonance broadening) and the collisions with the charged particles (Stark). Stark broadening of spectral lines as function of electron density ($N_e$) is conveniently tabulated, especially in the case of hydrogen lines[43]. In order to get the Stark width of the spectral line one needs to de-convolute the rest of "widths" from the measured line profile. For moderate-high pressure plasmas, collisions between neutral particles can contribute significantly to the line broadening and need to be considered.[44] Instrumental broadenings for different spectrometer gratings were measured using calibration lamps. The upper limits for Doppler broadening and Van der Waals (VdW) broadening are evaluated assuming that the gas temperature at its maximum approaches the electron temperature, i.e. $T_n = T_e$ , where $T_n$ is the temperature of neutrals. $T_n$ is considered equivalent to the gas temperature $T_g$. The upper limit for resonance broadening is evaluated assuming maximum density of hydrogen atoms, i.e. full dissociation of the $H_2$ comprising 10% of 500 Torr gas pressure. Relevant calculations of the resonance broadening and VdW broadening were detailed in Ref. 45. All broadening widths relevant the described experiments are summarized in Table 2. It shows that broadening due to neutral collisions and Doppler broadening are negligible.

Table 2. Instrumental, resonance, VdW, and Doppler broadening FWHMs of the Hα line used for diagnostics.

| $T_g$ (eV) | $\Delta_{Dopp}$(nm) | $\Delta_{VdW}$(nm) | $\Delta_{res}$(nm) | $\Delta_{instr}$(nm) |
|---|---|---|---|---|
| 0.3 | 0.027 | 0.076 | 0.01 | 0.7 (gr 150), 0.079-0.038 nm (gr 1200 ) |
| 0.6 | 0.04 | 0.096 | 0.01 | 0.7 (gr 150), 0.079-0.038 nm (gr 1200 ) |

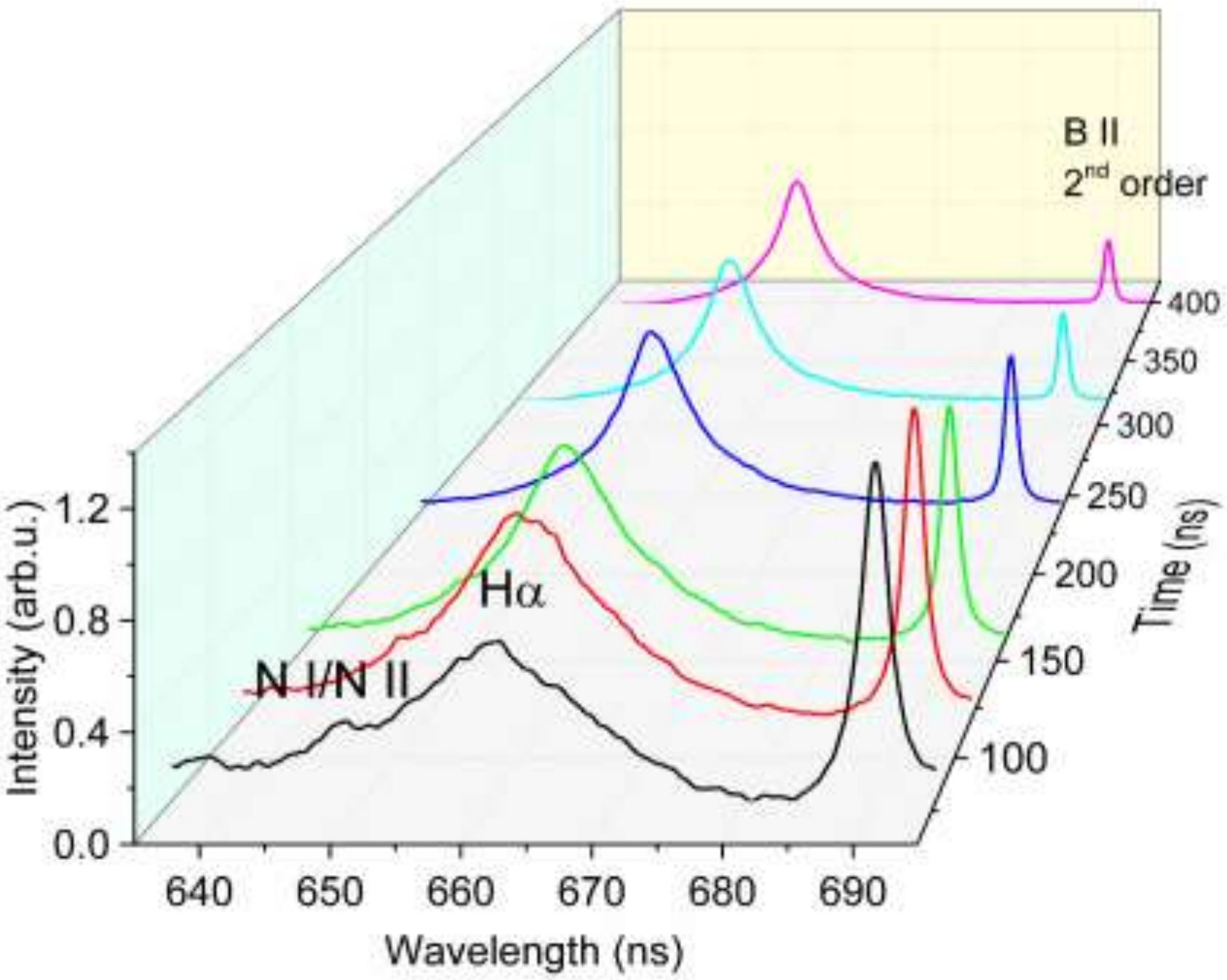

Figure 6. Temporal evolution of the $H_\alpha$ line profile through first 400 ns after the laser shot.

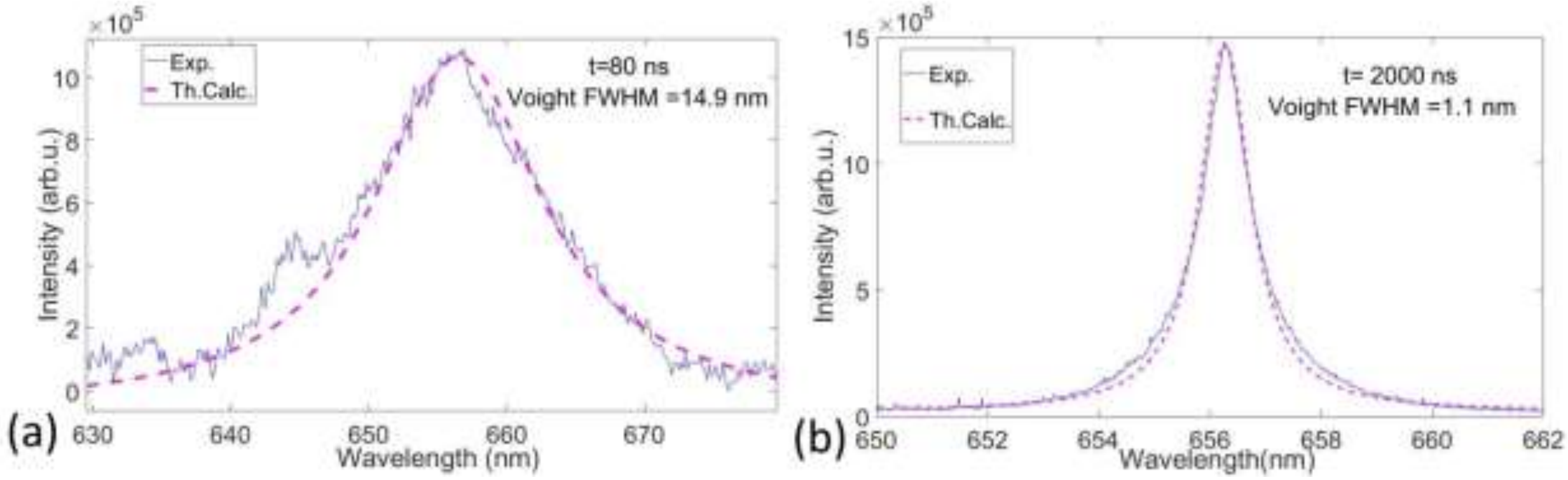

Figure 7. Examples of experimental and fitted $H_\alpha$ profiles obtained during B target ablation at P=500 Torr $N_2$+10% $H_2$ (a) t=80 ns, profile obtained with low resolution grating (150 gr/mm) (b) t=2000 ns, profile obtained with medium resolution grating (1200 gr/mm).

The experimental $H_\alpha$ line profiles are fitted with Voight profiles (i.e. convolution of Gaussian and Lorentzian), where the Gaussian contributions is due to instrumental and Doppler broadening and Lorentzian is due to VdW, resonance and Stark. In the current case, it is safe to assume the deconvolution of instrumental broadening from the experimental line one is left with Stark-broadened profile of the spectral line. Figure 6 shows the calculated Voight profiles compared with experimental $H_\alpha$ profiles captured at t=80 and 2000 ns. The full width-half maximum (FWHM) and full-width-half-area (FWHA) of hydrogen Balmer lines is tabulated as function of $N_e$ in the monumental work by Gonzales et al.[43], and was shown to depend very weekly on electron temperature $T_e$. In the same work, it is claimed that FWHA is more stable parameter, therefore we used the diagnosis maps of $H_\alpha$ FWHA for $\mu$=1, found in supplemental materials. It should be noted that calibration expressions for FWHM and FWHA of Balmer series were derived in the same work and are very convenient for the use[46] as opposed to diagnosis maps. However, we noticed that using the calibration expression (for $H_\alpha$ FWHA) yields $N_e$ larger by factor ~2.6 than the values obtained from diagnosis maps. Therefore, we have opted to use the diagnosis maps, as a more particular approach.

Figure 8 shows the temporal evolution of the electron temperature and the plasma density during the ablation for boron and BN targets in the $N_2$+10% $H_2$ ambient gas mixture at the pressure 500 Torr. Since the emission of $H_\beta$ was observed at t≥60 ns, we could not deduce the excitation temperature from these measurements at the earlier times. At the later times, the electron temperature varies between 1 to 0.3 eV. This implies that the condition of Eq. (2) is satisfied for the ablation of both boron and boron nitride targets with the plasma densities of $n_e$≥1.57 and 2.88 $\times 10^{13}$ $cm^{-3}$. With such high plasma densities reached at t<5 μs, the local thermal equilibrium is satisfied and therefore we can refer to the temperature measured from the ratio of hydrogen Balmer lines as electron temperature, rather than excitation temperature.

For both boron and boron nitride targets, the plasma densities are comparable between 80<t<700 ns (Figure 8). In the case of the BN target, the plasma density decays quicker than that for the boron target after 700 ns. Furthermore, the Balmer series emission for the BN target disappears after 1100 ns, while for the boron target, both the $H_\alpha$ and $H_\beta$ emission lines remain detectable up to t=5 μs. With the earlier detection of Balmer lines for the BN target than for the boron target, the electron temperature for the BN target shows also higher value around 0.6 eV at t=200 ns and goes down to 0.35 eV at t~500 ns. For the boron target, the measured electron temperature remains nearly constant around ~0.35 eV for the whole

period 300<t<1000 ns.

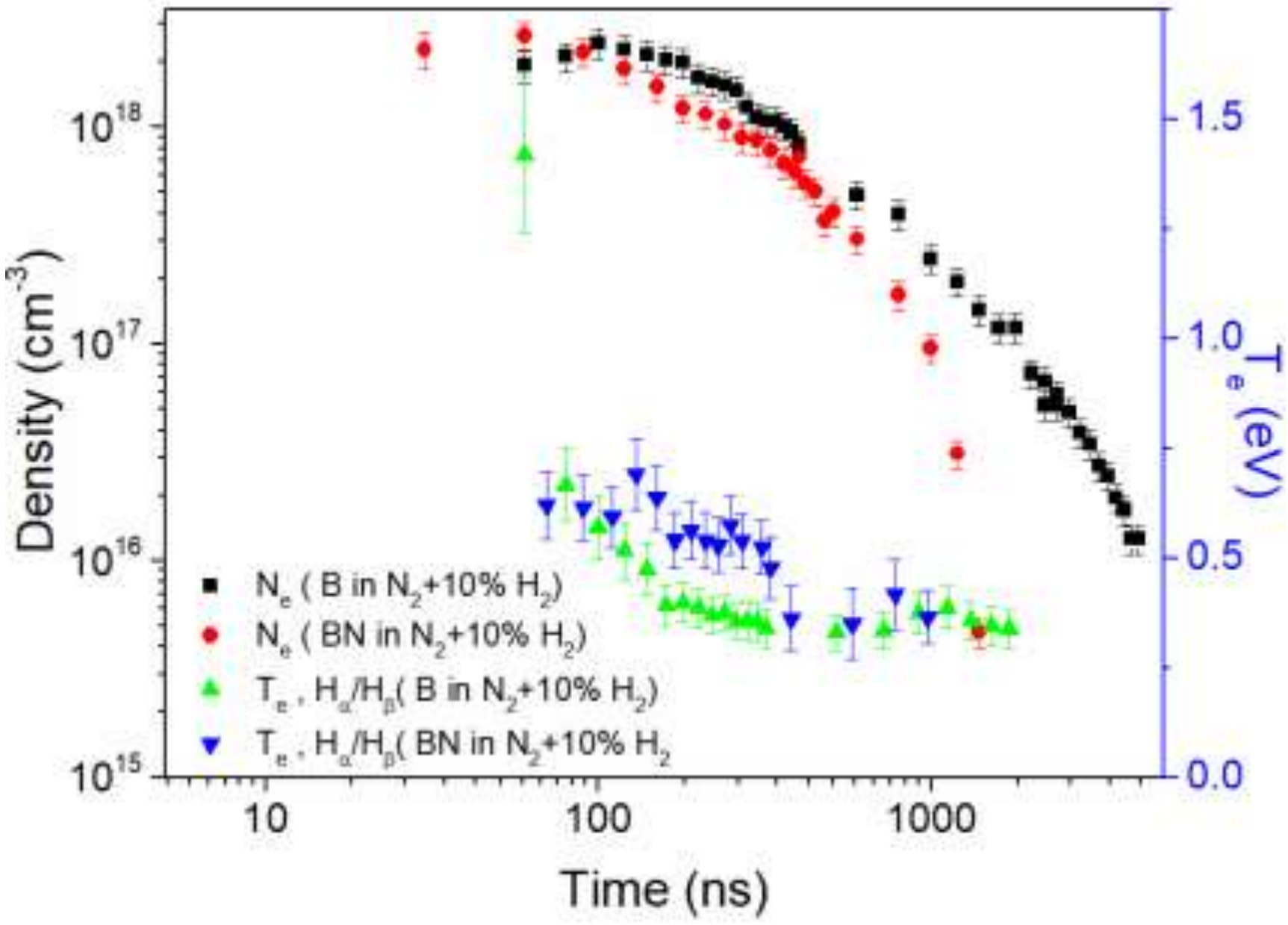

Figure 8. Temporal evolution of the Ne and Te during ablation of B and BN targets in 500 Torr of $N_2$+10%$H_2$.

Note that no significant differences in the line broadening/plasma density were found along the axis perpendicular to the target. The calculated densities and temperatures represent the averaged values across the whole plasma volume.

3.4 Quantification of ablated material

In order to evaluate the amount of material ablated per laser shot we have conducted an experiment in which each target was weighted before and after a 10 min ablation experiment. For each target, the experiment was repeated 6 times in vacuum (P=100 mTorr) and in 500 Torr $N_2$. We found no significant difference in the ablated weight of a target immersed in vacuum (P=100 mTorr) or high pressure (P=500 Torr). Table 3 summarizes the ablation experiments in terms of the duration, number of laser pulses per the experiment, ablated weight and corresponding density. For the boron target, we assumed that the ablated weight is comprised out of only B atoms. For the case of BN target, we assumed that the ablation is either by separate B and N atoms or by BN molecules (i.e. the ratio of ablated B and N is 1). The estimation of the ablation plume volume relies on the spatial imaging of the emission lines (e.g. the propagation of the emission line edge with respect to the target edge, see Appendix B) and images of the ablation plume profile. Using the spatial imaging we estimate that within the ~40 ns after the laser pulse, the ablation plume can approximated by a cylinder with a length of ≈ 0.2±0.03 cm and a diameter ≈ 0.1±0.03 cm, i.e. a plume volume of ≈1.6 nm$^{-3}$. This volume is used in order to estimate the number density of the ablated species at the very early times following the laser pulse.

Table 3. Determining the weight and number density ablated from each target per laser pulse. The results are generalized for P=100 mTorr and P=500 Torr (nitrogen ) cases, as no significant difference was found in the experiments

| Target | Ablation experiment duration (sec) | # of pulses | Total ablated mass (gr $^{-3}$) (P=100 mTorr and 500 Torr) | Ablated mass per shot (gr $^{-9}$) (P=100 mTorr and 500 Torr) | # of ablated B/ B and N atoms | Ablation Plume length (cm) at t=40 ns | Number density (cm$^{-3}$) of ablated species |
|---|---|---|---|---|---|---|---|
| B | 600 | 6000 | 0.15 ±0.069 | 25 | 1.764×10$^{16}$ | 0.2 | (1.123±0.643)×10$^{-19}$ |
| BN | 600 | 6000 | 0.29 ±0.057 | 48.333 | 3.25×10$^{16}$ | 0.2 | (2.068±0.876)×10$^{-19}$ |

3.5 Plasma characterization

The following section analyses the case of ablation of B target, at nitrogen pressure P=500 Torr. The measured plasma density and electron temperature in the case of BN target is very similar to the case of B target for 80<t<700 ns, therefore, within this time range the discussion can be generalized to BN case also. The laser ablation involves the evaporation of the target material, the formation of the plasma followed by the plasma decay. During these processes, the plasma density is determined by a trade of between the plasma generation, plasma expansion (for ablation to vacuum) and recombination. For the target ablation in collisional environments of the sub-atmospheric pressure ambient gas, the plasma generation and recombination govern the temporal evolution of the plasma density. An example of this evolution for boron target in nitrogen atmosphere (P=500 Torr) is shown in Figure 9. Here, the ionization degree (Figure 9a) is defined as:

$$Z = \frac{N_e(t)}{N_n(t=0) - N_e(t)} \ (4).$$

In addition to the ionization degree, collision frequencies shown in Figure 9a were calculated using measured $N_e$ and $T_e$. The definitions of these frequencies are summarized in Appendix A. Apparently, our measurements captured the end of the processes where the plasma formation dominates over the recombination. The ionization degree initially increases reaching its peak value of 9 % at 100 ns. This process is followed by the plasma decay when the ionization degree drops to >> 0.1%. During this plasma evolution, electrons are likely in thermal equilibrium between themselves as electron-electron collisions dominate over collisions of electrons with ions and neutrals. Because of a much higher density of the neutrals (Z<<100%), elastic collisions of electrons with neutrals dominate over electron-ion collisions. These electron-neutral collisions contribute to the heating of atoms and molecules in the ablation plume (Appendix A).

According to Figure 9, the plasma decay takes more than 2 microseconds. In the absence of electric field, the charge densities decay with time according to

$$\frac{dN_e}{dt} = -\beta N_e N_i, \quad (5)$$

where $\beta$ is the recombination coefficient. Here, we neglected the contribution of the plasma expansion in the reduction of the plasma density as it should not be a significant factor for the ablation in the surrounding gas (nitrogen or helium) The recombination coefficient  can be determined experimentally via plotting $N_e^{-1}$ as a function of the time (Figure 9a). From Figure 9a, we find $\beta \approx 1.65\times10^{-12}$ cm$^3$/s for $100<t<240$ ns, $\beta \approx 3.19\times10^{-12}$ cm$^3$/s for $260<t<400$ ns, and $\beta \approx 5.83\times10^{-12}$ cm$^3$/s for $600<t<1600$ ns.

The recombination can be governed by different mechanisms including radiative, dissociative and three-body processes, or their combination. It is instructive to determine which mechanism or mechanisms are dominant in the described ablation experiments. The radiative recombination ($A^+ + e^- \rightarrow A + h\nu$) coefficient, is given by[Error! Bookmark not defined.47]:

$$\beta_{rr} \approx 2.7\text{x}10^{-13} (T_e[eV])^{-\frac{3}{4}} \text{ cm}^3/\text{s} \quad (6)$$

Dissociative recombination ($A_2^+ + e^- \rightarrow A + A^*$) relies on the presence of molecular ions. However, our measurements did not record molecular ions in the emission spectrum (see Figure 10). Furthermore, the value of dissociative recombination coefficient $\beta_{dis}$ is ~$10^{-7}$ cm$^3$/s for temperature range from room to several thousand K[47 Error! Bookmark not defined.], decreasing as $\sqrt{T_e}$. Since in the present experiments, the measured plasma decay corresponds to $\beta$ ~$10^{-12}$ cm$^3$/s, the dissociative recombination is apparently not relevant for the ablation plume from boron and boron nitride targets.

Finally, three-body recombination ($A^+ + e^- + e^- \rightarrow A + e^-$) coefficient is given by:[47]

$$\beta_{crr} \approx 8.75\text{x}10^{-27} (T_e[eV])^{-\frac{9}{24}} n_e \text{ cm}^3/\text{s} \ (7).$$

For the conditions of our experiments, this amounts to ~ $10^{-8}$-$10^{-7}$ $cm^3/s$ which is also much larger than the experimental values of $\beta$ ~$10^{-12}$ $cm^3/s$.

Thus, the above analysis suggests that starting from t > 100 ns; the dominant recombination mechanism is radiative recombination. This may not be true for very early phases of the ablation (i.e. t≤ 100 ns), when ions recombine to produce neutrals and three-body recombination is the dominant mechanism. To demonstrate it we assume a simplified scenario in which neutrals are solely produced due to three-body recombination, so that the neutral density production rate is:

$$\frac{dn_n}{dt} = \beta_{crr} \times n_i \times n_e \quad (8).$$

Assuming that the plume plasma is quasi-neutral and taking the peak measured electron density we have $n_i = n_e = 1 \times 10^{18}\ cm^{-3}$ and $\beta_{crr} \approx 1.1\text{x}10^{-8} cm^3/s$ (for $T_e = 0.5\ eV$), corresponding to neutral production rate of $\frac{dn_n}{dt} \approx 10^{28}\ cm^{-3}/s$. With such a high rate a density of $n_n \sim 10^{21}\ cm^{-3}$ can be generated in period of 100 ns. This simplified estimation shows that fast recombination due to three-body recombination can generate large neutral density on scale of 100 ns, which is the delay between the observed peaks of ion and neutral emissions, during BN target ablation in vacuum (Fig.4a).

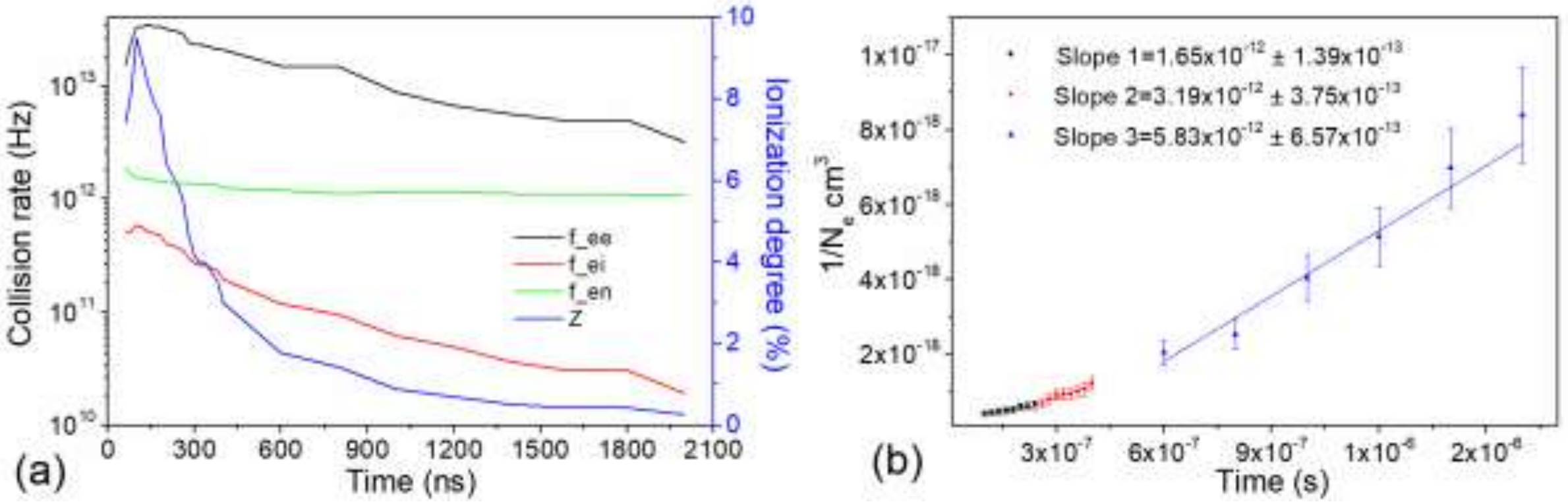

Figure 9. Ablation of B target in gas (P=500 Torr, 90 %$N_2$+10%$H_2$) (a) Collision rates and ionization degree as a function of time (b) plot of $1/N_e$ in a time range 100-2000 ns and linear fits of 3 time-segments of the curve.

**4.Discussion**

4.1 Chemical kinetics: BN target in vacuum and in He.

The following section discusses the emission dynamics during ablation of BN target, in order to understand the origins of BN and $B_2N$ molecules formation. For the ablation of BN targets in helium and vacuum, BN and $B_2N$ emission peak appear earlier then the emission of B and N neutrals (Fig. 4a). This result suggests that the BN molecules originated either directly from the solid target or are formed in the very early stage of the ablation. This hypothesis is also supported by the spatial location of the molecular emission (BN and $B_2N$), that appears directly adjacent to the target (Fig. 5a). The emission of BN appears at the same time as the emission from B and N ions (Fig. 4a). The Einstein coefficients ($A_{ki}$) for the examined ion and neutral lines of B and N are similar ($10^7$-$10^8$ $s^{-1}$) nevertheless, the emission peaks of neutrals are delayed by >100 ns with respect to their ion counterparts. For example: for the N I 746.8 nm and N II 500.5 nm lines, the coefficients $A_{ki}$=1.96e7 $s^{-1}$ and 1.14e8 $s^{-1}$, respectively.[38] They correspond to natural lifetimes of ~51 and ~ 9 ns, respectively. Therefore, the delay between emissions of N I and N II, which are simultaneously excited by collisions with energetic electrons should be ~40 ns. In a strongly collisional environment of the ablation-generated plasma, we can expect the effective lifetimes of the excited levels to be even shorter, and so should be their difference. In the experiment however, the measured delay between the peak emissions of N I and N II is ~100 ns. Hence, one can conclude that the neutral population reaches its peak density later than ions by approximately 60 ns. Two possible scenarios can be considered to rationalize the lack of emission by neutrals in the early stage: (1) vast majority of ablated atomic species are ions and molecules (2) neutrals are rapidly consumed in reaction that generates molecular species, before they can exhibit significant emission of their excited levels.

If scenario (1) is correct the ablated products should include BN and possibly $B_2N$ molecules and B and N ions. Fast collisional quenching in the high density area renders that effective lifetimes of the excited vibronic levels to be short: our experiment shows emission within tens of nanoseconds as opposed to typical molecular natural lifetimes on the order of microsecond. In such circumstances, the peak emissions by ions and molecules appear within first 100 ns (Fig. 4a, phase I). Recombination of ions creates neutrals, whose intensity (and density) peaks at t~250 ns (Fig. 4a, phase II). Following the analysis and estimations done in section 4.1, we can conclude that scenario (1) is plausible. The abundance of neutrals prompts a renewed formation of BN molecules, whose emission exhibits another peak at t~500 ns (Fig.4a, phase III). It has to be mentioned here that no correlation between BN and $B_2N$ emissions was observe at this time range. Moreover, there is no late increase in BN (0,0) emission during BN ablation in helium (Fig. 4c).

In scenario (2), it is hypothesized that the BN molecules are rapidly produced in the high density region which is formed immediately following the ablation near the target. In section 3.6, we estimated that in the case of BN target ablation at t=40 ns, there would be a volume of ≈1.6 $nm^{-3}$ occupied by B and N atoms or ions (here we assume that BN molecules are not emitted directly from the target). Then, the neutral density in this volume is n=2.07x$10^{25}$ $m^{-3}$. Since the optical emission of ions is already observed at

t=20 ns, let us consider that the half of the ablated species are ions and half are neutrals, i.e. the number density of B and N neutrals is $n_{n_B}=n_{n_N}=5.175\times10^{18}cm^{-3}$. Following Ref.[48] the rate coefficient $k$ for reaction N+B+M→BN+M was shown[36]**Error! Bookmark not defined.** to be $10^{-32}$ $cm^6s^{-1}$, where "M" represents all components in the system. In this case, the formation rate of BN: $\frac{dn_{BN}}{dt}=kn_Bn_Nn_M\approx5.54\times10^{24}$ $cm^{-3}s^{-1}$. Thus, within 40 ns after the ablation starts, the density of BN molecules would amount to ~$2.22\times10^{17}cm^{-3}$. If all these molecules are excited, their density should be sufficient to exhibit emission. The considered reaction rate does not occur under isochoric conditions and relies on the above simplifications. Therefore, the above value constitutes a rough estimate. Nevertheless, it demonstrates that the outlined scenario is not completely far-fetched.

Note that a high BN density would also promote the rapid formation of $B_2N$ molecules, via BN+B+M→$B_2N$+M, resulting in slightly delayed emission peak of $B_2N$. The data that is currently available does not allow us to completely evaluate the viability of this scenario. Passive diagnostics approach such as OES is inherently limited in regards of investigating the chronological appearance of species on short timescales of few nanoseconds. Therefore, one would need to apply an active diagnostic approach for measurement of existence and timing of appearance of B and N neutrals and BN molecular species.

To conclude this section, it appears that both above scenarios are plausible, based on rough estimations, although scenario (1) demands fewer assumptions and thus, seems more probable than scenario (2).

### 4.2 Chemical kinetics: BN target in $N_2$ gas.

The ablation of BN target in $N_2$ gas is quite different from any other ablation case investigated in this work. For example, unlike ablation in helium and vacuum, the BN ablation does not feature a peak of BN and $B_2N$ emission within the first 100 ns. First recognizable signals from both molecules appear at t=100 ns. The peak of BN (0,0) emission is reached at t~500 ns, which is similar to the local maximum of BN (0,0) emission obtained for BN target ablation in vacuum. This may be attributed to the BN formation by the B and N neutrals after both these species reach their density peaks (as manifested by peaks of optical emission, Fig.4d , t=100 ns).

### 4.3 Chemical kinetics: B target in $N_2$ gas.

For the case of the boron target ablation in nitrogen, no emission of BN (0,0) is observed until t=350 ns. The increase of the camera exposure to 100 ns did not help to detect BN emission. This is in contrast to the ablation of the BN target for which BN emission was observed right after the laser pulse with the camera exposure of 20 ns. A possible explanation of this result is that in order to form BN molecules, sufficient stocks of B and N species have to be built up in the plume. While boron is abundant from the target, the nitrogen atomic species must be formed via dissociation of $N_2$. It is generally acceptable that the dissociation of $N_2$ at moderate-to-high pressures (≥1.5 Torr) occurs mainly via vibrational excitation rather than by direct electron impact.[49,50] Therefore, one would expect to see the emission from excited $N_2$ molecules. However, this is not the case in the described experiments because no emission from $N_2$ or $N_2^+$ was observed (see Fig. 10). The dissociation energy of the nitrogen molecule is 9.8 eV. It may be that a population of high energy electrons capable to dissociate $N_2$ is large enough before 50 ns (i.e. $T_e$ is several eV), but not later, i.e. when we start measuring the temperature (≤ 1 eV) using line emission ratio (Eq. 1). The same may be applied for step dissociate processes which are likely to take place at collisional environment of the ablation plume ($\lambda_{mfp}$ << d, L). It is likely that using active diagnostics for electron temperature (Thomson scattering) would be useful to measure temperature before the emission from the necessary atomic lines appears. The reactions that do not involve dissociation of $N_2$, e.g. ($N_2$+B↔BN+N) have also been proposed.[51] However, for the best of our knowledge, there is no data available about this reaction and its rate. The viability of the dominance of this reaction may also be probed by using laser induced fluorescence (LIF) approach to measure $N_2$ and boron densities and their dynamic behavior.

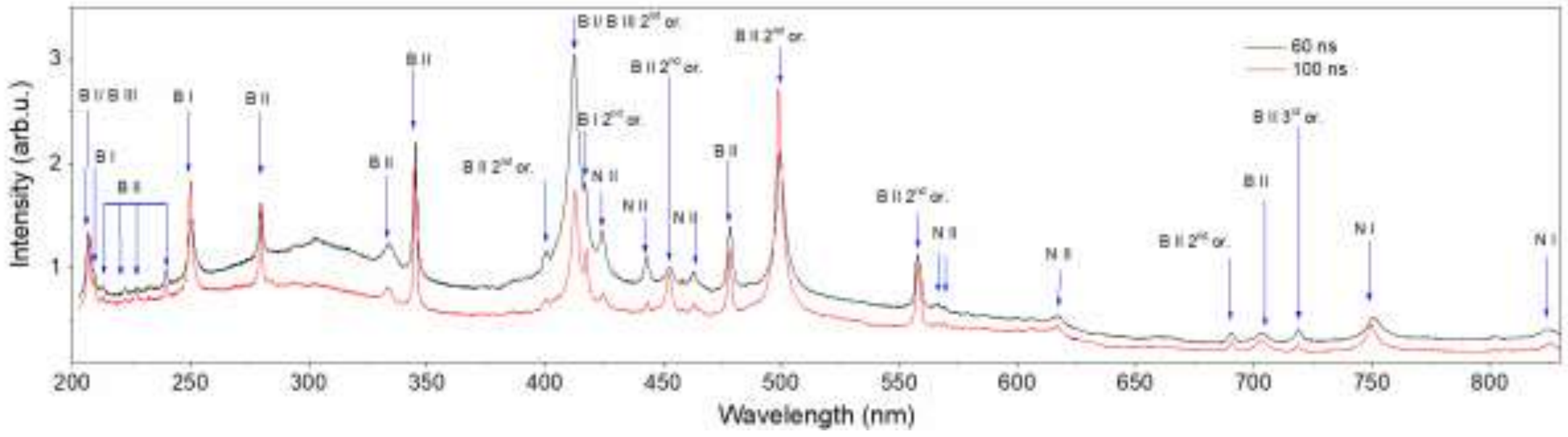

Figure 10. Broadband spectral emission in range 200-830 nm for B ablation in $N_2$. This spectrum is not corrected with the grating response function in order to preserve the readability of the figure. The apparent emission maximum at ~ 300 nm is because the spectrometer grating is blazed at 300 nm. No emission from $N_2$ or $N_2^+$ bands is observed. The camera gate time is 20 ns, the spectrum is averaged over 10 accumulations.

## 5. Conclusions

In this work, we presented the first integrated characterization of plasma properties (density and temperature) and chemical reactivity during laser ablation of boron and boron nitirde targets. The peak plasma density values are ~$10^{18}$ $cm^{-3}$ (remaining so for ~200 ns following the laser pulse) and measured peak electron temperature is ~ 1eV (decreasing to ~0.3-0.4 eV after). The presence of $B_2N$ molecules is manifested by molecular emission of these species when the ablated target is BN, disregarding the background gas or lack of thereof. It is suggested that the BN and $B_2N$ species are either emitted directly from the target or rapidly formed in the dense plasma region formed adjacent to the target during first tens of ns after the laser pulse. In order to decisively determine between these two scenarios an active diagnostics approach based on laser induced fluorescence could be used.

In the case of boron target ablation, the chemical kinetics in the ablation plume is different. Although atomic species emission behaves almost identically to the BN target case, the emission of BN molecules occurs much later (~300 ns), while no emission of $B_2N$ is observed. The delay in the emission and alleged formation of BN is apparently due to the necessity to generate the feedstock of N atoms, which in the case of boron ablation in $N_2$ atmosphere must come from dissociation of nitrogen molecules. It is hypothesized that the formation of BN depletes the atomic N feedstock and therefore, prevents a detectable production of $B_2N$ molecules in the plume.

Thus, the laser ablation of the BN target promotes the formation of a larger variety of BN-based molecular species including both BN and $B_2N$, than the ablation of the boron target in nitrogen atmosphere which generates only BN molecules. This result may have a practical implication for synthesis of BNNT in plasmas. Specifically, the use of a solid BN ablation target or powder should enable a more efficient generation of boron and nitrogen feedstock for the synthesis of BNNTs than the ablation of boron target or powder.

**Acknowledgements**

The authors are grateful to Dr. M. Schneider, Dr. Igor Kaganovich and Dr. Alexander Khrabry for fruitful discussion and advice. The experiments were supported by the US DOE, Office of Science, Basic Energy Sciences, Materials Sciences and Engineering Division.

## Conflicts of interest

There are no conflicts to declare.

**Appendix A: Calculation of collision frequencies and ionization degree**

electron-neutral collision rate[52] : $\nu_{en} = n_n \sigma_{en} v_{eth}$ (1),

where neutral density $n_n = \frac{P_g}{k_B T_n}$ , $\sigma_{en}$=$10^{-16}$cm$^2$ electron-neutral collision cross section, $v_{eth} = \sqrt{\frac{8k_B T_e}{\pi m_e}}$ thermal velocity of electrons.

Electron-ion collision rate: [52] $\nu_{ei} = 4.8 \times 10^{-8} n_i(cm^{-3}) \times ln\Delta \times T_i^{-\frac{3}{2}} \times \mu^{-\frac{1}{2}}$ (2) ,

where $ln\Delta = 7.47 + 1.5 \times \log(T_e(K) - 0.5 \times \log(N_e(cm^{-3}))$ **Error! Bookmark not defined.** [47] is the Coulomb logarithm (~4 for our conditions) and $\mu = \frac{m_i}{m_p}$ is the ratio of ion mas to that of proton.

electron-electron collision rate: [52] $\nu_{ee} = 2.9 \times 10^{-6} n_e(cm^{-3}) \times ln\Delta \times T_e^{-\frac{3}{2}}$(3) ,

Here we have assumed quasi-neutrality of the plasma, i.e. $n_i = n_e$ and that $T_i$ ≈4200 K (evaporation temperature of B). Let us briefly discuss the validity of this assumption. The rate of ion heating by neutrals can be evaluated via $\frac{dT_i}{dt} = (T_e - T_i) \times \frac{2m_e}{M} f_{ei}$ . For values $f_{ei} = 5.07 \times 10^{11} s^{-1}$ , $T_e \approx 1.06\ eV, T_{i0} \approx 0.026\ eV$ (cold background gas) and $T_{i0} \approx 0.36\ eV$ (ablated material) we get $\frac{dT_i}{dt}$=4.085x$10^7$ $\frac{eV}{s}$ and 2.75x$10^7$ $\frac{eV}{s}$ for the cases of cold background gas and hot ablated material, respectively. Using these heating rates one can see that ion temperature is very close to $T_e$ after only 20 ns.

**Appendix B: Spectral lines spatial evolution and hydrogen Balmer series appearance**

The time of flight of excited species can be evaluated via the imaging of species emission lines evolution. Figure A-1 shows a collage of images presenting the advancement of ion emission in spectral range 280-440 nm. The abscissa is the wavelength in nm and the ordinate is the distance from target edge. The presented images are collected during ablation of solid BN target at pressure P=100 mTorr. The velocity with which the front of ion emission proceeds is estimated to be ~2.8±0.1×$10^6$ cm/s (using the bright emission of NII lines at 392 and 393 nm and B II emission at 412.2 nm).

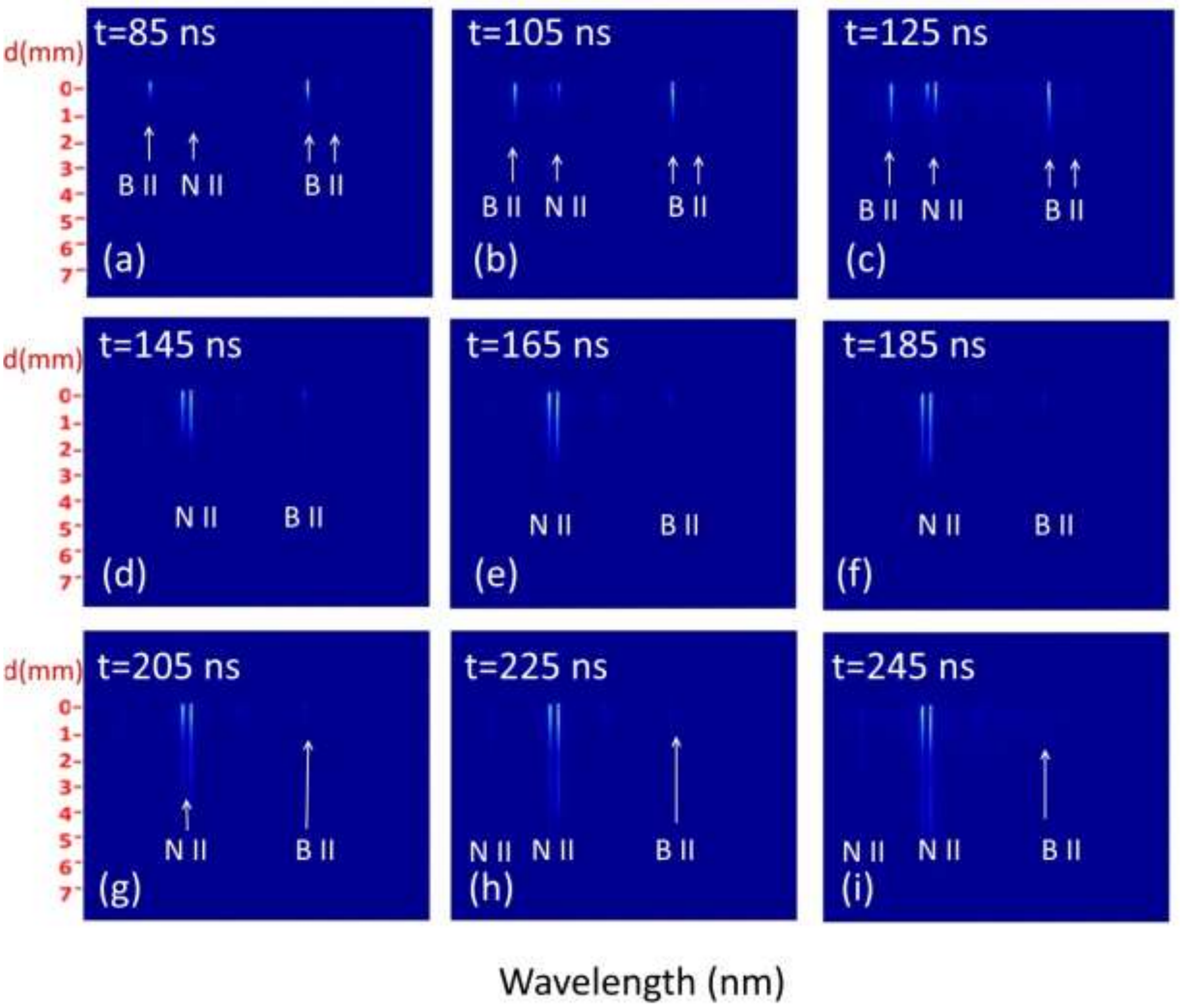

Figure A-1. Imaging the spectral lines spatial evolution during the early period following the laser shot at BN target at pressure P=100 mTorr.

Figure A-2 shows the broadband spectra in range 450-670 nm during ablation of B target in $N_2$+10% $H_2$ gas mixture, at pressure P=500 Torr. The observed lines indicated by species on the plot. As can be seen at the plot the peak of Balmer β line starts to be distinguishable at about t=100 ns and therefore this is when we can start "measuring" temperature, via line intensity ratio.

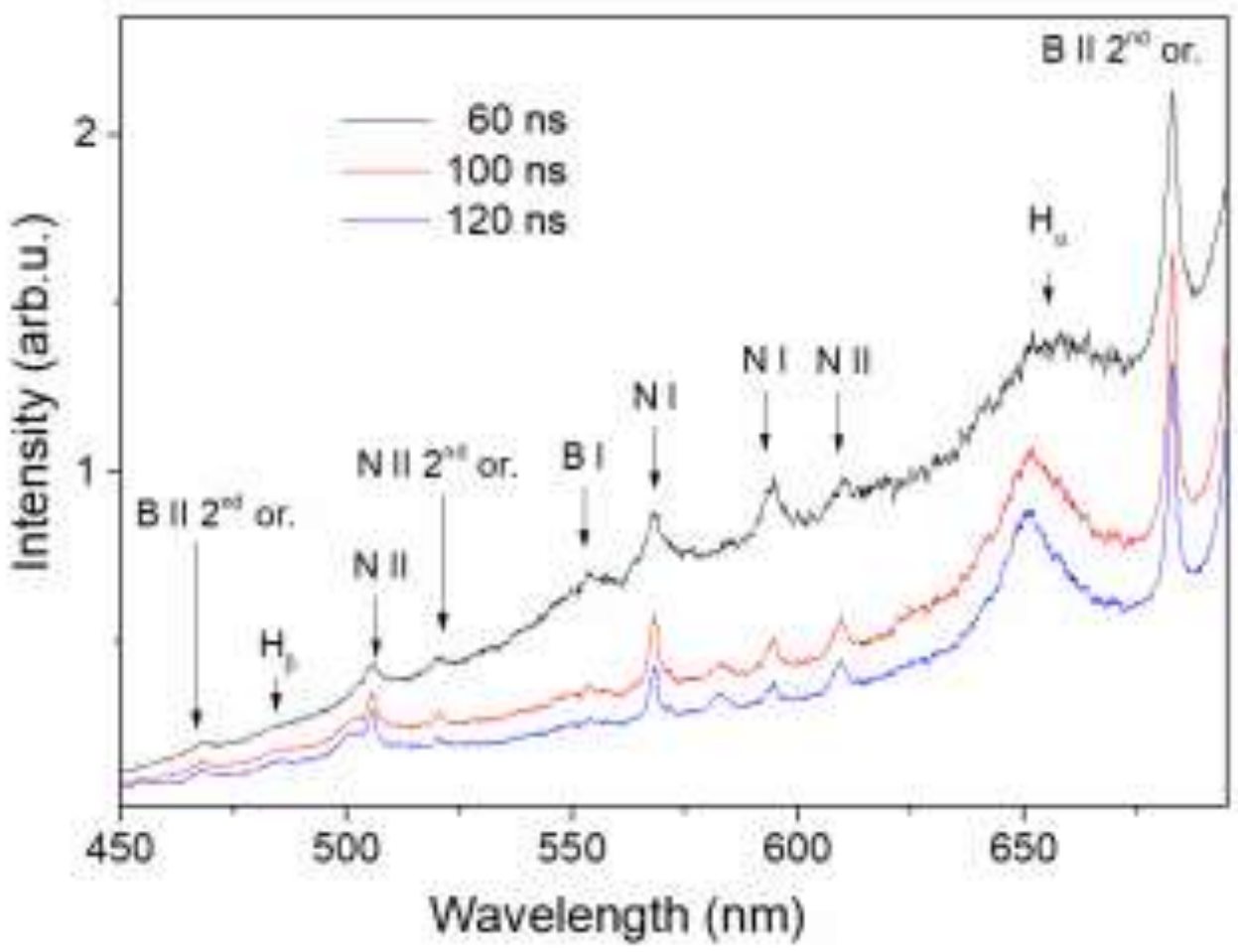

Figure A-2. Broadband emission featuring first appearance of hydrogen Balmer α and β lines, during ablation of B target in $N_2$+10% $H_2$ gas mixture. The spectrum was corrected for the grating response.

[9] A. Pierret, L. Schue, F. Fossard, J. Barjon, O. Ersen, S. Moldovan, F. Ducastelle and A. Loiseau, *"Optical and structural properties of facetted boron nitride nanotubes"*, European Microscopy Congress 2016: Proceedings, 476-477.

[10] R. Agrawal, A. Nietro, H. Chen, A. Agarwal, *"Nanoscale damping characteristics of boron nitride nanotubes and carbon nanotubes reinforced polymer composites"*, ACS Appl. Mat. Interfaces **5,** 12052 (2013).

[11] X. Hong, D. Wang, D.D.L. Chung, *" Strong viscous behavior discovered in nanotube mats, as observed in boron nitride nanotube mats"*, Composites Part B. **91**, 56 (2016).

[12] J.H. Kang, G. Sauti, C. Park, V. I. Yamakov, K.E. Wise, S. E. Lowther, C. C. Fay, S. A. Thibeault and R. G. Bryant, *"Multifunctional electroactive nanocomposites based on piezoelectric boron nitride nanotubes"*, ACS Nano **9**, 11942 (2016).

[13] L. H. Li and Y. Chen, *"Superhydrophobic properties of nonaligned boron nitride nanotube films"*, Langmuir **26**, 5135 (2010).

[14] P. Ahmad, M.U. Khandaker and Y. M. Amin, *"Synthesis of highly crystalline multilayers structures of BNNTs as a potential neutron sensing element"*, Ceramics International **41**, 4544 (2015).

[15] M. Ghazizadeh, J.E. Estevez and A.D. Kelkar, *"Boron nitride nanotubes for space radiation shielding"*, Int. J. Nano Studies and Tech **4**, 1 (2015).

[16] V. Raffa, C. Riggio, M.W. Smith, K.C. Jordan, W. Cao, A. Cuschieri, *"BNNT-mediated irreversible electroporation: it's potential on cancer cells"*, Technology in Cancer Research and Treatment **5**, 459 (2012).

[17] G. Ciofani, V. Raffa, J. Yu, Y. Chen, Y. Obata, S. Takeoka, A. Menciassi and A. Cuschieri , *"Boron nitride nanotubes: a novel vector for targeted magnetic drug delivery",* Curr. Nanosci. **5**, 33 (2009).

[18] J. Cumings and A. Zettl, *"Mass-production of boron nitride double-wall nanotubes and nanococoons"*, Chem. Phys. Lett. **316**, 211 (2000).

[19] Y. W. Yeh, Y. Raitses, B. E. Koel and N. Yao, *"Stable synthesis of few-layered boron nitride nanotubes by anodic arc discharge"*, Sci. Rep., **7**, 3075 (2017).

[20] R. Arenal, O. Stephan, J.-L. Cochon and A. Loiseau, "Root-growth mechanism for single-walled boron nitride nanotubes in laser vaporization technique", J. Am. Chem. Soc., **129**, 16183 (2007).

[21] D. P. Yu, X. S. Sun, C. S. Lee, I. Bello, S. T. Lee, H. D. Gu, K. M. Leung, G. W. Zhou, Z. F. Dong, and Z. Zhang, *"Synthesis of boron nitride nanotubes by means of excimer laser ablation at high temperature"*, Appl. Phys. Lett. **72**, 1966 (1998).

[22] L. C. Nistor, G. Epurescu, M. Dinescu and G. Dinescu, *"Boron nitride nano-structures produced by pulsed laser ablation in acetone"*, IOP Conf. Ser.: Mater. Sci. Eng. **15**, 012067 (2010).

[23] A. Fathalizadeh, T. Pham, W. Mickelson and A. Zettl, *"Scaled synthesis of boron nitride nanotubes, nanoribbons, and nanococoons using direct feedstock injection into an extended-pressure, inductively-coupled thermal plasma",* Nano Lett. **14**, 4881 (2014).

[24] K. S. Kim, C. T. Kingston, A. Hrdina, M. B. Jakubinek, J. Guan, M. Plunkett and B. Simard, *"Hydrogen-catalyzed, pilot-scale production of small-diameter boron nitride nanotubes and their macroscopic assemblies"*, ACS Nano **8**, 6211 (2014).